\documentclass{aa}
\usepackage{graphics}
\usepackage{txfonts}
\usepackage{natbib}
\bibpunct{(}{)}{;}{a}{}{,}

\begin{document}

\title
{Low accretion rates at the AGN cosmic downsizing epoch}
\subtitle{}

\titlerunning{AGN cosmic downsizing}

\author
{
A.\ Babi{\'c}\inst{1} \and
L.\ Miller\inst{1} \and
M.\ J.\ Jarvis\inst{2} \and
T.\ J.\ Turner\inst{3,4} \and
D.\ M.\ Alexander\inst{5} \and
S.\ M.\ Croom\inst{6}
}

\authorrunning{A.\ Babi{\'c} et al.\ }

\institute{Department of Physics, University of Oxford, 
Denys Wilkinson Building, Keble Road, Oxford OX1 3RH, U.K.
\and
Centre for Astrophysics Research, STRI,
University of Hertfordshire, Hatfield, Herts AL10 9AB, U.K.
\and
Department of Physics, University of Maryland Baltimore 
County, 1000 Hilltop Circle, Baltimore, MD 21250, U.S.A.
\and 
X-Ray Astrophysics Laboratory, NASA Goddard Space Flight 
Center, Greenbelt, Greenbelt, MD 20771, U.S.A.
\and
Department of Physics, Durham University, South Road, Durham 
DH1 3LE, U.K.
\and
School of Physics A28, University of Sydney, NSW 2006, 
Australia
}

\date{Received / Accepted}

\abstract 
{
X-ray surveys of Active Galactic Nuclei (AGN) indicate ``cosmic
downsizing'', with the comoving number density of high-luminosity
objects peaking at higher redshifts ($z\sim2$) than low-luminosity
AGN ($z<1$).
}
{
We test whether downsizing is caused by activity shifting towards
low-mass black holes accreting at near-Eddington rates, or by a
change in the average rate of accretion onto supermassive black
holes. We estimate the black hole masses and Eddington ratios of an
X-ray selected sample of AGN in the \textit{Chandra} Deep Field
South at  $z<1$, probing the epoch where AGN cosmic downsizing has
been reported. 
}
{
Black hole masses are estimated both from host galaxy stellar
masses, which are estimated from fitting to published optical and
near-infrared photometry, and from near-infrared luminosities,
applying established correlations between black hole mass and host
galaxy properties. Both methods give consistent results.
Comparison and calibration of possible redshift-dependent effects
is also made using published faint host galaxy velocity dispersion
measurements.
}
{
The Eddington ratios in our sample span the range $\sim 10^{-5} -
1$, with median $\log \left( L_\mathrm{bol} / L_\mathrm{Edd}
\right) = -2.87$, and with typical black hole masses
$M_{\mathrm{BH}} \sim 10^8$\,M$_\odot$.  The broad distribution of
Eddington ratios is consistent with that expected for AGN samples
at low and moderate luminosity.  We find no evidence that the CDF-S
AGN population is dominated by low-mass black holes accreting at
near-Eddington ratios and the results suggest that diminishing
accretion rates onto average-sized black holes are responsible for
the reported AGN downsizing at redshifts below unity. 
}
{}

\keywords{accretion; galaxies: active}

\maketitle

\section{Introduction}
One of the most significant recent discoveries in studying the
cosmological evolution of AGN has been the discovery that
lower-luminosity AGN ($L_{2- 10\,\mathrm{keV}}\sim 10^{41} - 10^{
43}$\,erg\,s$^{ -1}$) peak in their comoving space density at lower
redshift, $z\la1$, than higher-luminosity AGN ($L_{2-
10\,\mathrm{keV}} \sim 10^{45} - 10^{ 47}$\,erg\,s$^{ -1}$), which
peak at $z\sim2$
\citep{cowie,steffen,ueda,barger,giacconi,lafranca,hasinger,hopkins07}.
 
The implication of the phenomenon for black hole growth can be seen
from the emissivity of AGN in different luminosity ranges
\citep[Fig.\,5b]{hasinger}: at redshifts $z\gtrsim2$ the major
contribution to the total emissivity comes from high luminosity AGN
($L_{0.5 - 2\,\mathrm{keV}} \sim 10^{44} -10^{45}\,$erg\,s$^{-1}$),
while at $z\sim0$ the major contribution shifts to lower
luminosities ($L_{0.5 - 2\,\mathrm{keV}} \sim 10^{43} -
10^{44}\,$erg\,s$^{ -1}$), with a significant contribution from
low luminosity AGN ($L_{0.5 -2\,\mathrm{keV}} \sim 10^{42}- 10^{43
}\,$erg\,s$^{ -1}$). If we simply relate the emissivity to the
accretion rate onto black holes, it follows that a significant
contribution to the black hole growth has shifted from high
luminosity objects at high redshifts to low luminosity ones at low
redshifts.  This shift has been named `AGN cosmic downsizing', by
analogy with the phenomenon observed in the cosmic star formation
rate density in normal galaxies \citep[`cosmic downsizing',
e.g.][]{cowie96,bauer05,panter}.  There, the decline in star
formation rate at redshifts below $z\lesssim 2$ has been shown to
be due to star formation preferentially stopping in high mass
galaxies towards the present day, and the major contribution to the
star formation density at $z\sim 0$ arises from low mass systems.
The relationship of galaxy downsizing to AGN downsizing is unclear:
the phenomenon in AGN is observed in {\em luminosity}, and so an
interpretation of the phenomenon as being downsizing in the mass of
active black holes requires either an assumption of the
relationship between AGN luminosity and black hole mass as a
function of redshift or measurement of that relationship.
Supporting evidence that low-luminosity AGN at low redshift are
preferentially powered by black holes with lower masses than the
typical distribution of local black holes has been found when using
[OIII] emission as a measure of activity \citep{heckman}, although
when radio activity is measured instead, higher-mass black holes
seem more active locally \citep{best05}.

In this paper we present a determination of the distribution of
Eddington ratios for  an X-ray selected sample of AGN in the CDF-S,
at $z<1$. This redshift range is of particular interest, as it is
the epoch in which the total AGN emissivity is dominated by
moderate and low luminosity AGN --- the AGN downsizing epoch.  The
Eddington ratios are found by first estimating black hole masses
from photometrically-estimated stellar masses and luminosities.
These black holes masses are then compared with the  host galaxy
velocity dispersions from  \citet{vdw}. The Eddington ratio is then
calculated as $\lambda\equiv L_{\rm bol} / L_{\rm Edd}$ where
$L_{\rm bol}$ is the AGN bolometric luminosity inferred from the
observed X-ray luminosity and $L_{\rm Edd}$ is the Eddington
luminosity. We assume $H_0 =71$\,km\,s$^{ -1}$\,Mpc$^{ -1}$,
$\Omega_M = 0.3$ and $\Omega_{\Lambda} = 0.7$ throughout.

\section{The AGN sample}
The CDF-S has been the subject of a 1~Ms {\em Chandra} observation
reaching a flux limit of $4.5 \times 10^{ -16}\,$erg~cm$^{
-2}$~s$^{ -1}$ in the 2--10~keV band \citep{rosati}. At redshift
$z=1$ the flux limit corresponds to a hard X-ray luminosity of
$L_{2 - 10\,\mathrm{keV}} = 10^{ 42.4}$~erg~s$^{ -1}$, meaning that
the AGN population that dominates the downsizing, with $L_{2 -
10\,\mathrm{keV}} \sim 10^{ 41.5} - 10^{ 43.5}$\,erg\,s$^{ -1}$ at
$z < 1$, is observed in the survey.  Optical imaging and
identifications have been published by \citet{giacconi} with
optical spectroscopic follow-up by \citet[hereafter S04]{szokoly},
providing spectroscopic redshifts,  in addition to X-ray and
optical classifications.  The total number of X-ray sources in the
catalogue is 347, of which 251 are detected in the hard X-ray band.
Availability of the spectroscopic (S04) and photometric redshifts
\citep{zheng,mainieri} together with the deep X-ray exposure has
allowed the X-ray spectroscopic analysis published by
\citet{tozzi}. For bright sources, these authors fitted  the photon
spectral index $\Gamma$ and the absorbing hydrogen column density
$N_\mathrm{H}$ simultaneously, while for fainter sources the photon
spectral index was fixed at the mean value $\langle
\Gamma\rangle=1.8$. The $N_\mathrm{H}$ distribution was found to
have a log-normal shape, peaking at $\log (N_\mathrm{H}/\mathrm{
cm}^{-2}) = 23.1$.  From the fitted spectra, absorption-corrected
X-ray luminosities were estimated and have been made publicly
available by \citeauthor{tozzi}.  There are 321 objects classified
by their absorption-corrected total X-ray luminosity as AGN ($L_{X}
> 10^{ 41}$~erg~s$^{ -1}$), and the completeness of the
spectroscopic (X-ray) catalogue with respect to the photometric one
is $\sim 99\%$. 

Deep BVIz optical photometry in CDF-S is provided by the GOODS
survey, from ACS/HST imaging \citep{giavalisco} with sensitivity
limits 27.8,~27.8,~27.1,~26.6 ({\em AB} magnitudes, 10$\sigma$
limits, in $0\farcs2$ aperture). The deepest R-band data available
(R$_{AB}\sim26.7$) are included in the 1~Ms catalogue
\citep{giacconi}. In the near-infrared (NIR), \citet{olsen} have
re-reduced the ESO Imaging Survey data \citep{arnouts}, with the
observations reaching median limiting magnitudes J$_{AB}\sim23.1$,
K$_{AB}\sim22.2$ ($5\sigma$ limits, in $2''$ aperture). In the
infrared, the Spitzer Wide-area InfraRed Extragalactic survey
(SWIRE) provides observations in four IRAC bands:
3.6,~4.5,~5.8,~8.0$\,\mu$m \citep{surace}. The 90\% completeness
limits for SWIRE Data Release 2 are $m_{AB}=
(21.0,20.96,19.84,19.53)$. CDF-S is also one of the fields covered
by the COMBO-17 survey, with photometry in 12 medium-band and 5
broad-band filters \citep{wolf}. The COMBO-17 magnitude limits are
24.7,~25.0,~24.5,~25.6,~23.0 in U,~B,~V,~R,~I bands respectively
({\em AB} magnitudes, $10\sigma$ limits, $\sim1\farcs5$ aperture),
while limiting magnitudes in the 12 medium-band filters are in the
range $m_{AB}=22.0-23.9$.

Recently the deepest CDF-S optical to infrared public data has been
compiled into a uniform photometric catalogue of NIR-selected
objects by the GOODS-MUSIC (MUltiwavelength Southern Infrared
Catalog) project \citep{grazian}. The catalogue is based on imaging
in BVIz from ACS/HST, JHK$_s$ from ISAAC/VLT, U band from the
2.2ESO and VLT-VIMOS, and 4 IRAC/Spitzer bands. The authors have
developed a `PSF-matching' (point spread function) algorithm, which
allows finding precise object colours between ground-based and
space-based images.

In this paper we define and analyse two samples drawn from the
above data.  We limit the redshift to $z<1$, as we are interested
in probing the epoch for which AGN downsizing has been reported
\citep{barger}. Given the X-ray survey flux limits, we are
guaranteed to probe the AGN population predominantly responsible
for the inferred downsizing.  We include all $z<1$ objects X-ray
classified as AGN ($L_X>10^{41}\,\mathrm{erg}\,\mathrm{s}^{-1}$).
There are 150 objects in the \citeauthor{tozzi} catalogue that
satisfy our redshift and X-ray luminosity criteria.

Objects in our `spectroscopic' sample are those selected to be AGN
with secure redshift measurement, secure optical identifications
and drawn from a catalogue with uniform photometry.  Analysis of
the sample uses the BRVIzJHK$_s$ and 3.6$\,\mu$m photometry from
the GOODS-MUSIC catalogue. The catalogue has been matched to the
\citeauthor{tozzi} AGN, with a 4 arcsec radius search, and each
match has been manually checked. For objects with photometric
redshifts where identification was not secure, the nearest of any
multiple matches is kept. The selection is restricted to objects
with detections in at least 4 bands. The resulting catalogue
comprises 83 objects, from which 55 satisfy our `spectroscopic'
criteria, i.e. have secure optical identification and spectroscopic
redshift determination (redshift quality flag $3.0$ in
\citeauthor{szokoly} and \citeauthor{tozzi} catalogues).

The second sample analysed will contain the remaining objects
without a secure identification or spectroscopic redshift. To
obtain coverage of as many objects as possible, it is supplemented
with a combined catalogue consisting of the COMBO-17 BVRI
broad-band photometry, SofI JHK$_s$ \citep{olsen,moy} and IRAC
3.6$\mu\,$m \citep{surace} photometry, and the photometric
redshifts from \citet{zheng}.  The catalogues are matched to the
\citeauthor{tozzi} AGN, with a 2 arcsec search radius, keeping the
nearest of any multiple matches. We find matches for 41 more
objects with detections in at least 4 bands. The `photometric'
sample now comprises 69 objects.  The main purpose of creating this
`photometric' sample is to check for possible bias that might be
caused by restricting the sample to objects with secure optical
identifications and spectroscopic redshifts. This will be described
in more detail later. 

In summary, our spectral-template-fitting catalogues have in total
124 objects, i.e. 83\% of the AGN in CDFS with spectroscopic or
optical photometric redshifts $z<1$ \citet{szokoly,zheng}. Of
these, 55 comprise the `spectroscopic' and 69 the `photometric'
sample.

Absorption-corrected X-ray luminosities are adopted from the
\citet{tozzi} catalogue. Our analysis includes four additional AGN
from the RDCS1252-29 field which also have published measurements
of host galaxy velocity dispersion \citep{vdw} and for which we
have obtained new X-ray flux measurements from the {\em Chandra}
archival images \citep{rosati04}. The new fluxes are given in the
Table~\ref{table1}, and they correspond to absorption-corrected
fluxes assuming the mean CDF-S intrinsic absorption hydrogen column
density $\log(N_\mathrm{H}/\mathrm{cm}^{-2})=23.1$ and power-law
spectrum with the photon index $\Gamma=1.8$ \citep{tozzi}. Galactic
absorption hydrogen column density in the direction of the CL~1252
cluster is taken to be $N_\mathrm{H}=5.95\times10^{20}$cm$^{-2}$.
\begin{table}
\resizebox{\linewidth}{!}{
\begin{tabular}[h]{cccc}
\hline
\hline
ID & $\alpha$ & $\delta$ & $F_{2-10\,\mathrm{keV}}$  \\
& (J2000)  & (J2000)  & ($10^{-16}$~erg~cm$^{-2}$~s$^{-1}$) \\
\hline
CL 1252-1 & 12 52 45.8899 & -29 29 04.5780 & $15.6\pm 3.9$  \\
CL 1252-3 & 12 52 42.4793 & -29 27 03.5892 & $14.2\pm 3.6$  \\
CL 1252-5 & 12 52 58.5202 & -29 28 39.5256 & $5.83\pm 2.88$ \\
CL 1252-7 & 12 53 03.6396 & -29 27 42.5916 & $11.3\pm 3.3$  \\
\hline
\end{tabular}
}
\caption{
Absorption-corrected X-ray fluxes of four CL~1252 cluster
galaxies, measured from {\em Chandra} archival images.
Fluxes were calculated using the galactic absorption
hydrogen column density $N_\mathrm{H} = 5.95 \times 10^{
20}$cm$^{ -2}$ in the direction of the CL~1252 cluster,
mean CDF-S intrinsic absorption $\log(N_\mathrm{H}/
\mathrm{cm}^{ -2}) = 23.1$, power-law spectrum with the
photon index $\Gamma = 1.8$ \citep{tozzi}, and are given
in the observed 2--10~keV band. The optical counterpart
coordinates are listed.  Match of the X-ray sources to
the optical is within $2''$.
\label{table1}}
\end{table}

The aim is to find Eddington ratios for our AGN sample, for which
we need the black hole masses and bolometric luminosities.
Absorption-corrected hard X-ray luminosities together with
bolometric corrections will give the best available estimate of
bolometric luminosity. Black hole masses are more difficult to
estimate: in this paper we infer masses from available host galaxy
data and use established (at low redshift) correlations of
supermassive black hole masses and host galaxy properties. The deep
multi-wavelength data available for the field allows measurements
of the host galaxy properties that are needed for this approach.

\section{Black hole mass estimates}
\subsection{Host galaxy stellar mass and luminosity}
We have investigated two routes for estimating black hole masses
for our AGN: (1) finding stellar masses for the galaxies and
applying a relation between galaxy stellar mass and black hole mass
\citep{ferrarese06}; (2) finding the evolution- and $k$-corrected
K-band absolute magnitude and using the well-established
correlation between black hole mass and K-band bulge luminosity
\citep{marconi03}. Both are indirect methods, and using two
different methods should help demonstrate the robustness of the
general approach, i.e. finding the supermassive black hole masses
from galaxy properties.

In method (1) galaxy stellar masses are found by fitting spectral
templates to the measured spectral energy distributions (SEDs). SED
fitting requires precise relative magnitudes (colours), but for
inferring galaxy stellar mass the absolute normalisation of the SED
is also required. Furthermore, different catalogues give magnitudes
measured in apertures of differing diameter, and from images with
different seeing. To be as close as possible to a uniform
catalogue, we use the {\em SExtractor} \citep{bertin} automatic
aperture (Kron-like) magnitudes, which provide the most precise
total magnitudes for galaxies. The flux loss at faint magnitudes is
estimated to be $\sim 10\%$ with automatic magnitudes, compared to
$\sim 70\%$ with isophotal or $\sim 20\%$ with corrected isophotal
magnitudes. We use photometry in the  BRVIzJHK$_s$ and 3.6$\,\mu$m
bands. The longer wavelength IRAC bands are discarded because of
possible contamination by dust-reprocessed AGN light. The IRAC
3.6$\,\mu$m band is included, as it will be close to the rest-frame
near-infrared bands for many of our objects, and NIR light is
crucial for obtaining the correct stellar mass of the galaxy. U
photometry is not used, again because of possible non-stellar
continuum contributions. If there is further contamination with the
rest-frame UV AGN light, we expect to obtain a bad fit to the
template galaxy spectra: the possibility of a scenario where a
reasonable fit and non-stellar continuum conspire to indicate a
younger stellar population is discussed later. 

We use two different sets of templates from stellar population
synthesis models: \citet[hereafter BC]{bc93} and \citet[hereafter
M05]{m05}, to test the uncertainties coming from this part of the
method. For both sets we allow different types of templates,
variation in age of the stellar population and reddening. M05
templates include instantaneous bursts (SSP), exponentially
declining ($\tau=0.1,~0.3,~1,~2$~Gyr), truncated
($t_\mathrm{trunc}=0.1,~0.3,~1,~2$~Gyr) and constant star formation
histories; allowed metallicities are
$Z_\odot/5,~Z_\odot/2,~Z_\odot,~2Z_\odot$; reddening is varied in
the range $0\leq A_V\leq 3$; initial mass function (IMF) is
\citet{salpeter}. BC templates include instantaneous burst,
exponential ($\tau=1,~2,~3,~15,~30$~Gyr) and constant star
formation; IMF is \citet{millerscalo}, metallicities are allowed to
vary. The reddening law is \citet{calzetti} for both sets of
templates.

The redshifts are fixed to the spectroscopic values (S04).  For
fitting and stellar mass estimation we use a modified version of
the {\em Hyper-Z} code \citep{bolzonella}. Both sets of templates
produce similar stellar mass estimates. 

Black hole mass and the galaxy stellar mass are related using the
correlation involving the dynamical galaxy mass from
\citet{ferrarese06}:
\begin{equation}
  \log\left(\frac{M_\mathrm{BH}}{
  \mathrm{M}_\odot}\right) 
  = (8.47\pm0.08)+(0.91\pm0.11)
  \log\left(\frac{M_\mathrm{dyn}}{
  10^{11.3}\mathrm{M}_\odot}\right) ,
\label{eqdynsig}
\end{equation}
We discuss the validity of using stellar mass as a proxy for the
dynamical mass in the next section.

For the second method, we note that finding absolute K-band
magnitudes at higher redshifts would involve applying
$k$-corrections and, more importantly, significant evolutionary
corrections to the measured K-band magnitudes. To make this process
less arbitrary, we can utilise the rest-frame absolute K-band
magnitudes, which are found as a byproduct of the galaxy template
SED fitting procedure in the first method, and already contain the
evolutionary and $k$-corrections. We estimate the black hole masses
by applying the K-band bulge luminosity--black hole mass relation
\citep{marconi03}
\begin{equation}
  \log M_\mathrm{BH} 
  = (8.21\pm0.07)+(1.13\pm0.12)
  (\log L_{K,\mathrm{bul}} -10.9) ,
\label{eqmlkbul}
\end{equation}
where the mass and the luminosity are expressed in solar units.

The galaxy mass estimates from photometry rely on the galaxy light
not being contaminated by the AGN light. As already noted, we do
not use the infrared bands beyond 3.6$\,\mu$m as they are likely to
be affected by a detectable contribution of dust-reprocessed light.
There are further indications that we are indeed dealing with the
galaxy light only, i.e. that there is no significant non-stellar
continuum component in the considered bands. For 112 (118 for the
K-band luminosity method) out of 124 objects in our input catalogue
we are able to obtain a reasonable fit (in terms of $\chi^2$) with
a galaxy template, and the shape of the probability distribution
for stellar mass and K-band magnitude is such that it allows us to
measure their values and errors with reasonable confidence.
Furthermore, contamination in blue bands would make an object
appear younger, and have a lower stellar mass estimate, compared to
a pure galaxy-light object.  On the other hand, contamination in
the near-infrared bands would boost the NIR light, making the mass
estimate from the K-band higher.  Thus, we would see a large
difference in estimates from the stellar mass method and
K-magnitude method. 

The fitting procedure does not unambiguously determine the
morphological class of the object, but it does provide the most
likely star formation history for the galaxy and the age of the
stellar population. For more than half of the objects we find that
the age of the population is larger (often several times) than the
characteristic timescale for the last star-formation episode. About
20\% of the objects have the estimated age smaller than the
characteristic timescale for the galaxy template, and thus appear
to be star-forming. They do not show a large discrepancy in mass
estimate from the two methods, leading us to conclude that this is
not due to AGN-light contamination. Low or non-existing star
formation rate in most objects leads us to believe that it is
reasonable to approximate bulge light with total light, as the
inference is that these are early-type galaxies.  When applying the
relation (\ref{eqmlkbul}) we assume that this approximation holds:
the true bulge fraction would need to be significantly less to have
any large effect on our conclusions.  In any case, we apply an
empirical calibration in the next section that should, on average,
factor out this uncertainty.

\subsection{Comparison with velocity dispersion measurements}
This indirect way of determining black hole masses from
correlations with galaxy stellar properties results in black hole
masses with uncertainties of the order of the intrinsic scatter in
the correlations.  But in addition, systematic offsets could be
introduced if there is redshift evolution in those relationships,
given the relatively high median redshift for our sample, the
indirect way of arriving at the final result and the expectation
that these relations should evolve with cosmic epoch.  One of the
best-established proxies for black hole mass at low redshift is the
bulge velocity dispersion. Its possible redshift evolution has
received much attention \citep{woo06,treu}, although not yet with
definitive results.  Here, we take the view that this is the best
established relation, with low intrinsic scatter, and unknown
redshift evolution, and in this section we calibrate our other two
black hole mass estimators ($M_\mathrm{dyn}$ and $K_\mathrm{bul}$)
onto the same scale of black hole mass at $z\sim 1$ that is
obtained from the velocity dispersion.  This `re-calibration' is
performed using a sample of high redshift galaxies with known
velocity dispersion measurements.  Then, we can either assume that
the $M_\mathrm{BH}-\sigma$ relation is assumed not to evolve in
redshift range $z=0-1$, or we can apply the evolution in
$M_\mathrm{BH}-\sigma$ obtained by \citet{woo06} and \citet{treu}.
In terms of the results we obtain in section\,\ref{sec4} it turns
out that the former assumption is the more conservative (see the
later discussion on this point).  The shifts in black hole mass
introduced by this process are relatively small.

To perform the re-calibration, we use the sample of \citet{vdw},
who have made spectroscopic velocity dispersion measurements for 25
non-AGN galaxies and 4 AGN host galaxies in the CDFS field, of
which 23 have early-type morphology.  They also measured 5 non-AGN
and 4 AGN in the RDCS1252-29 field. This galaxy/AGN sample covers
the redshift range $0.62 < z < 1.13$, with median redshift 0.97,
and allows us to verify the applicability of our method and
re-calibrate the galaxy--black hole relations that we are going to
use (zero-points only, given the small size of the sample). There
are at least three reasons for re-calibrating the relations, coming
from the uncertainties associated with a higher-redshift sample:
stellar mass estimates are less reliable with increasing redshift
\citep{vdw06}; the reports on evolution in relations between the
galaxy properties and central black hole masses vary significantly
--- from reports of no evolution to redshift evolution so strong
that it would have to turn off by $z \la 1$ to be consistent with
other constraints on AGN and supermassive black hole populations
\citep{hopkins06c,woo06,treu,mclure06,peng06}; finally, we are
measuring the total galaxy light, but the local relations are given
in terms of bulge properties. It is important to emphasise that the
majority of the calibration sample consists of galaxies which have
no known AGN, and the assumption is the properties of the galaxies
are representative of the AGN hosts.

We match the available photometry in the field with the catalogue
of \citeauthor{vdw}, and find stellar mass estimates for those
galaxies, using both previously described methods. 

Galaxies in the sample have effective radii measurements and
dynamical mass estimates $M_\mathrm{dyn}=\alpha R_e\sigma^2 / G$,
with $\alpha=5$ \citep{vdw}.  We compare the reported dynamical
masses and our stellar mass estimates and find excellent agreement
{\em on average} (mean offset at the level of a few percent),
although with a large scatter.  Thus we proceed with using the
stellar mass estimate as a proxy for dynamical galaxy mass, keeping
in mind that the results are to be interpreted in an average sense,
with uncertainties for individual objects being large.

\begin{figure}[t]
\centering
\resizebox{0.95\linewidth}{!}{
\rotatebox{-90}{
\includegraphics{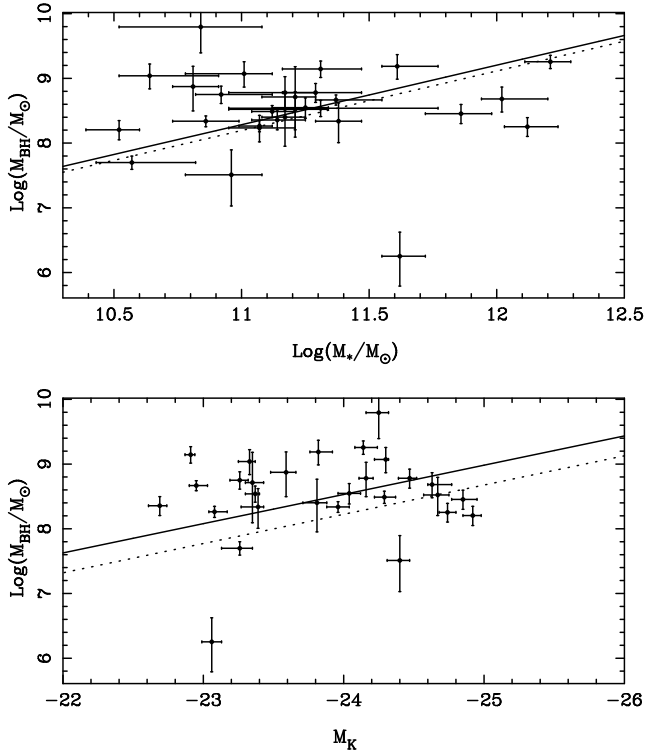}
}}
\caption{
Calibration of our two black hole mass estimate methods
using the measured velocity dispersions in CDF-S field,
for a sample of mainly normal galaxies (with 4 AGN in the
sample). 
a) Stellar mass method: black hole masses are calculated
from the measured velocity dispersions using relation
(\ref{eqmsigma}) and compared to stellar mass estimates. 
b) K-band magnitude method: black hole masses are again
calculated from the measured velocity dispersions using
relation (\ref{eqmsigma}) and compared to absolute K-band
magnitudes.  Dotted lines are the nominal zero-redshift
relations (\ref{eqdynsig},~\ref{eqmlkbul}). Solid lines
are shifted by the amount required to make the black hole
masses from measured velocity dispersions and model
quantities agree on average. 
\label{vdwfig}}
\end{figure}

We can now estimate black hole masses for the \citeauthor{vdw}
galaxies from the measured velocity dispersions using the
well-established (at zero redshift) relation between the black hole
mass and velocity dispersion \citep{ferrarese06}:
\begin{equation}
  \log\left(\frac{M_\mathrm{BH}}{
  \mathrm{M}_\odot}\right) 
  = (8.48\pm0.07)+(4.41\pm0.43)
  \log\left(\frac{\sigma_*}{
  224\mathrm{km s}^{-1}}\right) ,
  \label{eqmsigma}
\end{equation}
with a rms of $\sim$0.3 dex in $\log M_\mathrm{BH}$.  These black
hole mass estimates may then be compared with our mass estimates
using the same method as for our AGN hosts, namely applying
relations\,(\ref{eqdynsig}) and (\ref{eqmlkbul}) to stellar mass
and rest-frame, evolution-corrected K-band luminosities.  This
comparison tests for any differential evolution between the black
hole mass estimators and also allows us to place our black hole
mass estimates onto a common scale defined by the zero-redshift
$M_{\mathrm{BH}}-\sigma$ relation, but does not of course take into
account any evolution in that relation, which will be discussed
later.  Fig.~\ref{vdwfig} shows the results of this comparison.
The differences between the two stellar population synthesis models
are smaller than the uncertainties within one model, and we plot
results from M05 models only.  Top and bottom panels show the
stellar mass method and K-band luminosity method results,
respectively. K-band luminosity errors are smaller than stellar
mass errors because the rest-frame K-band luminosity is well
constrained by the 3.6$\,\mu$m measurements, whereas the variation
in the total mass yielded by different templates can be
significant.  Errors on plotted quantities are estimated by
identifying a change $\Delta\chi^2 = 1$ in the fits.  Dotted lines
are the nominal relations (\ref{eqdynsig},~\ref{eqmlkbul}), and the
solid lines are modified relations needed to obtain agreement
between our model predictions and measured velocity dispersions on
average. The zero-point shifts in the $M_\mathrm{BH}-M_*$ and
$M_\mathrm{BH}-L_K$ relations correspond to black hole mass shifts
$\Delta\log M_\mathrm{BH}=0.09$ and $0.31$ for the stellar mass and
K-band luminosity methods, respectively.  We again assume that the
host galaxy velocity dispersion is close to that of the bulge
velocity dispersion, as also assumed by \citet{woo06} and \citet{treu},
justified by the observation that velocity dispersion depends only
weakly on measurement aperture for early-type galaxies
\citep{jorgensen}.  As already noted by \citet{vdw06}, we find that
stellar population synthesis models typically underestimate the
velocity dispersions somewhat. This topic is beyond the scope of
this paper \citep[see][]{vdw06,drory,rettura}, and for our purposes
we simply rescale our mass estimates to obtain, on average, a match
to the measured velocity dispersions.

As noted above, we have thus far assumed that locally measured
relations (\ref{eqdynsig},~\ref{eqmlkbul},~\ref{eqmsigma}) hold at
higher redshifts, and have simply shifted the black hole masses by
$\Delta\log M_\mathrm{BH}=0.09$ and $0.31$ for the stellar mass and
K-magnitude methods, respectively to obtain agreement with
\citet{vdw} measurements. This way, we have factorised out any
redshift-related uncertainties in the stellar light from the host
galaxies.  Given that the shift that has been applied is relatively
small, the difference in median redshift between the AGN
`spectroscopic' (median redshift 0.7) and \citeauthor{vdw} (median
redshift 0.97) samples should be of little significance.

There is however experimental evidence that the black hole
mass--velocity dispersion relation (\ref{eqmsigma}) itself might
have redshift evolution, in the sense that at higher redshift,
galaxies with the same velocity dispersion host more massive black
holes than locally \citep{woo06,treu}. \citet{woo06} report a shift
in the black hole mass by $\Delta\log M_\mathrm{BH}=0.57\pm0.11$ at
$z=0.36$, similar to that reported by \citet{treu}.  At high
redshift, \citet{shields03} have reported no evolution in the
$M_{\rm BH}-\sigma$ relation whereas \citet{shields06} do find
evidence for evolution.  If evolution does occur, the bootstrapping
to the zero-redshift $M_{\rm BH}-\sigma$ relationship that we have
carried out would result in black hole masses being underestimated,
and the Eddington ratios that we estimate in the following section
to be overestimated.

\section{Eddington ratios in CDF-S}\label{sec4}
\subsection{Analysis of the `spectroscopic' sample}
\begin{figure}
\centering
\resizebox{7cm}{!}{
\rotatebox{-90}{
\includegraphics{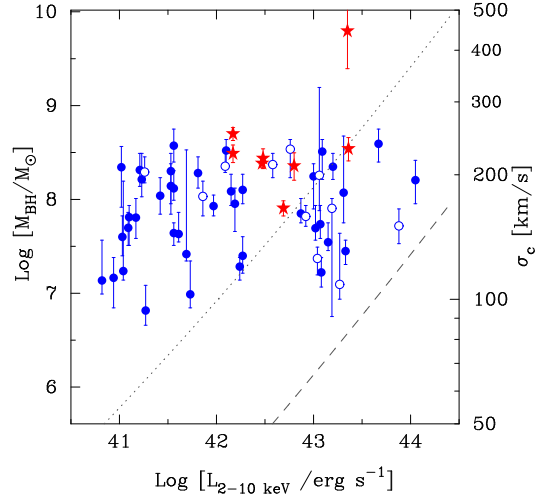}
}}
\caption{
Black hole masses and hard X-ray luminosities of the
`spectroscopic' sample of AGN in the CDF-S. Blue open
circles: black hole masses estimated using the stellar
mass method; red stars: black hole masses calculated from
measured host velocity dispersions for the AGN in
\citet{vdw}. Solid squares mark objects which show
presence of broad lines in optical spectra.  Lines
(dotted and dashed) denote lines of constant Eddington
ratio of 0.01 and 1.0, respectively. Redshift evolution
in the $M_\mathrm{BH} - \sigma$ relation
\citep[e.g.][]{woo06}, would cause a shift in all points
to higher mass by $\Delta \log M_\mathrm{BH} \sim 0.57$. 
\label{lxsigfig}}
\end{figure}

Having estimated the black hole masses for the sample, and given
the measured hard X-ray luminosities, we can now estimate Eddington
ratios (the ratio of an object's total luminosity and its Eddington
luminosity corresponding to its black hole mass,
$\lambda=L_\mathrm{bol}/L_\mathrm{ Edd}$). Bolometric luminosities
are obtained by applying the \citet{marconi} bolometric correction
to the absorption-corrected hard X-ray luminosities.  Black hole
mass estimates are described in the previous section. In summary,
they are calculated using the black hole--galaxy correlations, but
with zero-points shifted to obtain agreement with the available
high-redshift velocity dispersion measurements or dynamical mass
estimates. The absorption-corrected hard X-ray luminosities and
black hole mass estimates for our sample are shown in
Fig.~\ref{lxsigfig}, together with lines of constant Eddington
ratio of 0.01 (dotted line) and 1.0 (dashed line). 

Also shown on Fig.~\ref{lxsigfig} are black hole masses estimated
directly from the \citet{vdw} velocity dispersion measurements,
assuming the zero-redshift relation (\ref{eqmsigma}), for the 8 AGN
with such measurements.  It can be seen that these yield a similar
distribution to the stellar mass estimates, as expected from the
previous section.  This provides independent evidence that
systematic contamination from AGN light in the photometric mass
estimation is not a significant cause of bias.

\begin{figure}[bt]
\centering
\resizebox{\linewidth}{!}{
  \rotatebox{-90}{
  \includegraphics{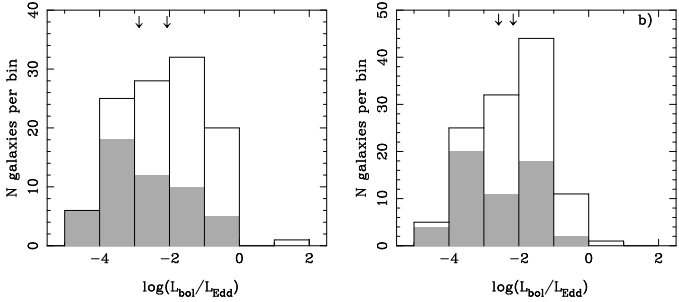}
  }
}
\caption{
Distribution of Eddington ratios for the `spectroscopic'
sample (shaded histogram) using (a) the stellar mass
method, (b) the K-magnitude method. Median Eddington
ratios are $\log\lambda =-2.87$ and $-2.58$, respectively
and the corresponding values are marked with arrows. The
solid line histogram shows the distribution of Eddington
ratios for the combined `spectroscopic' and `photometric'
sample.  Median Eddington ratios for the combined
distributions are $\log\lambda =-2.07$ and $-2.16$
(marked with arrows). Given the additional uncertainties
in the `photometric' sample, the joint sample is expected
to be broader than the distribution for the
`spectroscopic' sample only. The conclusions are largely
unaffected by the choice of sample. 
\label{eddhist}}
\end{figure}

Finally, we combine our bolometric luminosities and black hole
masses into Eddington ratio estimates. Fig.~\ref{eddhist} (shaded
histograms) shows the distribution of Eddington ratios for our
sample, using both stellar mass and K-magnitude methods. Median
(mean) values for the two methods are $\log\lambda=-2.76$~(-2.87)
and $-2.64$~(-2.58), respectively. 

\subsection{Analysis of the `photometric' sample}
Initially we have a hard X-ray selected sample, but the reduction
to the secure redshift and optical--X-ray identification subsample
could potentially be biasing us to brighter, and thus likely more
massive objects (as they are more likely to have spectroscopic
redshift measurement). To test for possible biases of our
`spectroscopic' subsample, we use the photometric redshift
estimates in CDF-S \citep{zheng}, which are available for
$\sim99\%$ of the X-ray detected sources, with the average redshift
accuracy $\sim 8\%$.  We supplement the non-secure sample with a
combination of photometric catalogues in the field, resulting in
photometric data for $\sim83\%$ of the whole $z<1$ X-ray selected
sample.  The combined sample colours are less precise than those
from the uniform GOODS-MUSIC catalogue.  In addition, neither
optical identifications nor the photometric redshifts are certain
for all objects, so there are likely to be several objects with
erroneous Eddington ratio estimates.  This makes the catalogue less
reliable compared to the `spectroscopic' one, but still the best
estimate we can make for the objects missing in our `spectroscopic'
sample. We proceed to find stellar mass estimates for the objects.
Fits with a well-defined minimum in $\chi^2$ are found for 61 out
of 66 objects. For broad-line AGN (18 and 11 objects in
`photometric' and `spectroscopic' samples, respectively) stellar
masses are more likely to be incorrect as the photometry could have
a significant contribution from the central object, but we do not
remove them from the sample. 
\begin{figure}[bt]
\centering
\resizebox{\linewidth}{!}{
\rotatebox{-90}{
\includegraphics{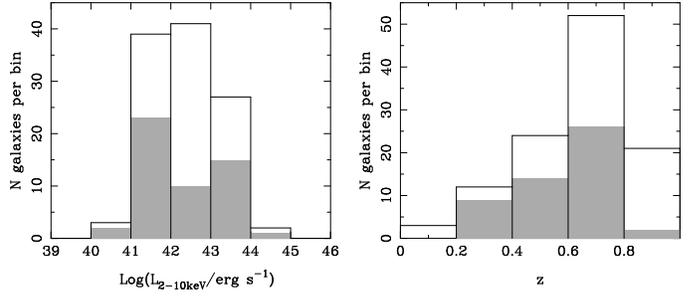}
}
}
\caption{
Distributions of absorption-corrected hard X-ray
luminosities (left panel) and redshifts (right panel) for
the `spectroscopic' sample (shaded histogram) and for
the combined `spectroscopic' and `photometric' sample
(open histogram). There is a shift to lower X-ray
luminosities and to higher redshifts for the
`photometric' sample.
\label{lxzhist}}
\end{figure}

We find that the distribution of Eddington ratios for all objects
(solid line histogram in Fig.~\ref{eddhist}), which includes the
`spectroscopic' sample (shaded histogram in Fig.~\ref{eddhist}) and
the `photometric' sample, follows closely the distribution of the
`spectroscopic' sample, convincing us that our subsample is not
biased towards optically bright objects with high black hole masses
and low Eddington ratios.  Fig.~\ref{lxzhist} compares the
luminosity and redshift distributions of the `spectroscopic' and
`photometric' samples.  Slightly higher redshifts (mean $\Delta
z\sim0.1$) make the `photometric' sample fainter than the optical
spectroscopy limits but those galaxies are not systematically less
massive.

\section{Discussion}
\subsection{Distributions of Eddington ratio}
The analysis presented here indicates that X-ray selected AGN at
the `cosmic downsizing epoch' in fact show a wide range in
Eddington ratio, with a median value $\log\lambda \sim -2.8$.  We
now discuss how this result compares with other determinations of
the distribution of AGN Eddington ratios in the literature.

At low redshift, \citet{panessa} find a similarly broad range of
Eddington ratio for local Seyfert galaxies selected on the basis of
optical spectra. They measured  nuclear X-ray fluxes for the
galaxies, and find an active nucleus in all but 4 out of 47 Seyfert
galaxies. The sample covers the bolometric luminosity range
$L_\mathrm{bol}\sim 10^{41}-10^{44}$erg~s$^{-1}$, very similar to
the one for our sample, but extending to lower luminosities because
of the ability to resolve the nucleus in X-rays. The black hole
mass range (estimated mostly from the $M_{\mathrm{BH}}-\sigma$
relation) is almost identical to our sample at $z \sim 0.7$.  The
estimated Eddington ratios for this sample again have a broad
distribution, in the range $\log\lambda=-1$ to $-7$, and median at
$\lambda \sim 0.01$ (type 1 AGN) or $\sim 10^{-3}$ (all AGN).

However, for bright ($L_\mathrm{bol}\sim 10^{ 44} - 10^{
47}$erg~s$^{ -1}$) local AGN, \citet{woo02} find Eddington ratios
typically in a narrower range 0.001--0.1, while at $z \sim 0.7$ the
ratios are somewhat higher, 0.01--1.  For optically selected AGN at
$0.3 < z < 4$, with bolometric luminosities in the range
$L_\mathrm{bol}\sim10^{ 45} - 10^{ 47}$erg~s$^{ -1}$, the Eddington
ratios have a narrower distribution (0.1--1), and no apparent
redshift dependence \citep{kollmeier}. The most obvious difference
with these samples is the different bolometric luminosity range,
even at $z < 1$. 

\begin{figure}
\centering
\resizebox{\linewidth}{!}{
\rotatebox{-90}{
\includegraphics{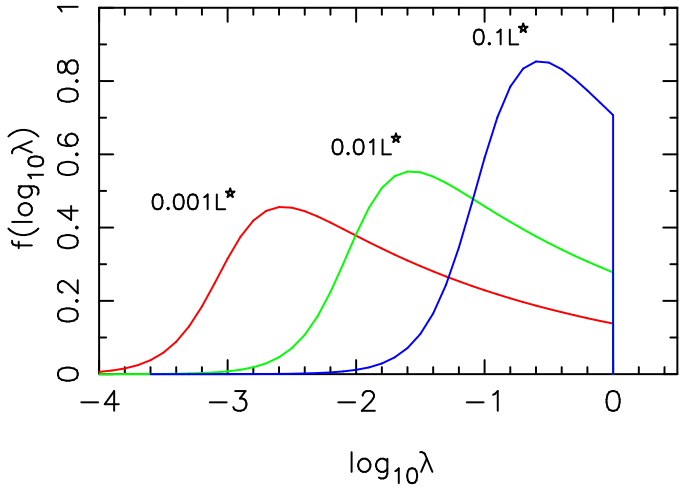}
}}
\caption{
Expected distributions of Eddington ratio $\lambda$ in
the simple illustrative model described in the text.
Three `samples' are assumed drawn from the active black
hole mass function of \citet{greene}, with a uniform
intrinsic distribution of $\log(\lambda)$, at
luminosities  $L=0.001 L^\star$ (red curve), $0.01
L^\star$ (green) and $0.1 L^\star$ (blue).
\label{examplefig}}
\end{figure}

The variety of reports on Eddington ratio mean value and width of
distribution can be understood from the following simple
illustration of selection effects.  Fig.\,\ref{examplefig} shows
the distribution of Eddington ratio we would expect to observe for
samples selected at three differing luminosities.  In this
illustration, the black hole mass function has been taken to be the
double power-law form for `active' black holes of \citet{greene},
the {\em intrinsic} distribution of Eddington ratios is taken to be
uniform in $\log\lambda$ over the range $-5 < \log\lambda < 0$, and
expected distributions for three choices of bolometric luminosity
are shown: $L=0.001 L^\star$, $0.01 L^\star$, $0.1 L^\star$, where
$L^\star$ is the Eddington luminosity for black holes at the
`break' in the double power-law mass function (\citealt{greene}
give the black hole mass at the `break' as $M_{\mathrm{BH}}^\star =
10^{7.32}$\,M$_\odot$ at $z=0$). 

The illustration demonstrates that, while drawn from the same
(unremarkable) {\em intrinsic} distribution,  the observed
Eddington ratio distribution is expected to be strongly dependent
on the luminosities probed by an AGN sample: selecting low
luminosity objects results in a wide Eddington ratio distribution
with a low mean value, whereas selecting high luminosity objects
results in a narrow distribution with a high mean Eddington ratio. 

Additionally, the observed distribution of $\log\lambda$ is a
function not only of the form of the intrinsic distribution but
also of the shape of the black hole mass function and the
luminosity selection and range covered by the sample.  Further
selection effects may also arise, if there are systematic
variations in AGN SEDs as a function of $\lambda$, as indicated by
\citet{boroson} and implied by the difference in Eddington ratios
typically seen for type\,1 versus type\,2 AGN
\citep[e.g.][]{panessa}.

\subsection{Cosmic downsizing}
Similar considerations to those above lead us to conclude that it
is necessary to understand the distribution of black hole masses in
AGN samples before physical interpretations may be made of the
phenomenon of `cosmic downsizing'.  The most striking result from
the analysis presented here is that the AGN responsible for the
peak in space density at $z \la 1$ at moderate AGN luminosities
cannot be described as being low-mass black holes accreting at high
rates.  This argues against the simplest interpretation of `cosmic
downsizing': the phenomenon is {\em not} due to an increasing
dominance of low-mass black holes, the typical black hole mass in
the CDF-S AGN sample is $\sim 10^8$\,M$_\odot$.

Downsizing thus appears to have a rather more complex origin.  The
typically low Eddington ratios we find could be consistent with the
model of \citet{hopkins06b}, where low-level AGN activity is
fuelled by stochastic accretion of cold gas, and dominates the AGN
population from $z=0$ up to redshifts $z\sim1$,  and at higher
redshifts merger-driven AGN fuelling might be dominant.  This could
be a generic expectation of hierarchical structure formation:
merger-driven bright phases of AGN activity exist in high redshift
universe, possibly alongside the less bright activity powered by
stochastic accretion. At low redshifts, once the mergers become
less frequent, stochastic accretion becomes the dominant mode. 

More generally, however, if black hole growth is coeval with galaxy
growth, then we expect the mean accretion rate onto galaxies, and
hence their associated black holes, to decrease with cosmic time
\citep{miller}.  Indeed, decrease in mean accretion rate must
occur, since the integrated AGN luminosity density decreases with
cosmic time at $z \la 2$, yet we do not expect the integrated mass
in black holes to decrease substantially with time (rather, the
latter should {\em increase} as black holes continue to grow).  Our
results suggest that the AGN cosmic downsizing at $z \la 1$ is not
a symptom of `anti-hierarchical' behaviour, but in fact may be a
reflection of the process of the dying-down of cosmic accretion and
a shift in the typical luminosity of massive black holes to lower
values.  Whether this trend continues to still lower redshift is
still an open question: the \citet{panessa} black hole mass and
Eddington ratio distribution appear similar to the CDF-S AGN of
comparable luminosity, yet it seems that lower-luminosity type\,2
AGN at low redshift are dominated by lower mass black holes with
moderate Eddington ratios \citep{heckman}.

Finally, we note that for a sample of high-redshift (median
$z=2.2$) sub-mm--selected galaxies in the {\em Chandra} Deep Field
North,  \citet{borys} have found that, if  their black holes are
assumed to be radiating at the Eddington limit, the black hole
masses in those high-redshift star-forming galaxies are 1-2 orders
of magnitude smaller than in galaxies of comparable mass in the
local universe. Given the high redshift of the sample, and the
discussion of evolution in mean accretion rate above, it may well
be that typical Eddington ratios are higher than for our sample.
But interestingly, if we assume that the effect they see is
actually from the sub-Eddington accretion rates similar to the ones
we report here ($\sim 0.01 L_\mathrm{Edd}$), this would bring the
stellar mass--black hole mass relation for high-redshift
star-forming galaxies into agreement with the local one.  Reliable
black hole mass estimates at higher redshifts would be a key
advance in understanding this area.

\section{Conclusions}
We have estimated Eddington ratios for two samples of hard X-ray
selected AGN in CDF-S with median redshift $z=0.7$.  The primary
`spectroscopic' sample has secure redshifts and optical
identifications and spans the bolometric luminosity range
$L_\mathrm{bol}\sim10^{41}-10^{45}$erg~s$^{-1}$. The majority of
the sources are radiating at low Eddington ratios in the range
$\lambda\sim1-10^{-5}$, with the median $\log\lambda=-2.87$. A
larger sample, based on photometric redshifts, has fewer selection
effects, but larger uncertainties related to optical identification
and Eddington ratio estimates: this fainter sample has Eddington
ratios that span the same range, and the median for the whole
sample is $\log\lambda=-2.07$. Black hole masses are in the range
$10^{5}- 10^{9}\,M_\odot$, with the distribution peaking at
$\sim10^{8}\,M_\odot$. It is likely that fainter X-ray flux limits
would reveal even more sources radiating at low Eddington ratios. 

We have discussed how the broad distribution of Eddington ratios
arises because of the relatively low luminosities probed by the
sample, and that in general observed distributions are strongly
dependent on the selected luminosity range of AGN samples.  Based
on the estimated Eddington ratios and black hole masses for the
CDF-S AGN, we argue that diminishing accretion rates onto
average-mass supermassive black holes \citep{miller} are the
underlying cause of the observed cosmic AGN downsizing at
$z\sim0.7$, contrary to an interpretation in which most of the
activity occurs in rapidly-growing low-mass black holes. 

\acknowledgements{
We thank Claudia Maraston for discussions on galaxy evolution
models and for providing galaxy templates, Micol Bolzonella for the
modified version of {\em Hyper-Z} code, Emanuelle Daddi for the
stellar mass macro, David Bonfield for the catalogue matching code.
AB acknowledges support from the Clarendon and Waverly Funds. DMA
thanks the Royal Society for support. MJJ thanks Research Councils
UK for support.
}

\bibliographystyle{aa}  
\bibliography{mygreatbibliography}
\label{lastpage}

\end{document}